\documentclass[twocolumn,english,nofootinbib]{revtex4-1}
\usepackage[T1]{fontenc}
\usepackage[latin9]{inputenc}
\usepackage[a4paper]{geometry}
\usepackage{amsmath}
\usepackage{amssymb}
\usepackage{cancel}
\usepackage{stmaryrd}
\usepackage[dvipsnames]{xcolor}

\makeatletter
\usepackage{hyperref}
\usepackage{slashed}
\usepackage{shuffle}

\makeatother

\usepackage{babel}
\begin{document}
\title{One-loop off-shell amplitudes from classical equations of motion}
\author{Humberto Gomez$^{a,b}$}
\author{Renann Lipinski Jusinskas$^{c}$}
\author{Cristhiam Lopez-Arcos$^{d}$}
\author{Alexander Quintero V\'elez$^{d}$}
\affiliation{$^{a}$ Department of Mathematical Sciences, Durham University, ~\\
Stockton Road, DH1 3LE Durham, UK}
\affiliation{$^{b}$ Facultad de Ciencias Basicas, Universidad Santiago de Cali,~\\
Calle 5 $N^\circ$ 62-00 Barrio Pampalinda, Cali, Valle, Colombia}
\affiliation{$^{c}$ Institute of Physics of the Czech Academy of Sciences \& CEICO
~\\
Na Slovance 2, 18221 Prague, Czech Republic}
\affiliation{$^{d}$ Escuela de Matem\'{a}ticas, Universidad Nacional de Colombia, \\ Sede Medell\'{i}n, Carrera 65 $\#$ 59A--110, Medell\'{i}n, Colombia}

\

\begin{abstract}
In this letter we present a recursive method for computing one-loop off-shell
integrands in colored quantum field theories. First, we generalize the
perturbiner method by recasting the multiparticle currents as generators
of off-shell tree level amplitudes. After, by taking advantage of
the underlying color structure, we define a consistent sewing procedure
to iteratively compute the one-loop integrands. When gauge symmetries
are involved, the whole procedure is extended to multiparticle solutions
involving ghosts, which can then be accounted for in the full loop
computation. Since the required input here is equations of motion and gauge symmetry, our framework naturally extends to one-loop computations in certain non-Lagrangian field theories.
\end{abstract}
\maketitle

\section{Introduction}

Scattering amplitudes are central objects of study in quantum field
theory. More than convenient physical observables, they are deeply
rooted in the very way we intuit particle interactions. And nothing
captures this statement more clearly than Feynman diagrams and their
beautiful simplicity.

We soon learn, however, that Feynman diagrams are far from being the
most efficient way of computing scattering amplitudes. There has been an impressive progress over the years in tree- and loop-level computations. Most of these developments
involve so-called \textit{on-shell methods} (see e.g. \cite{Dixon:1996wi,Bern:2007dw,Roiban2011ScatteringAI,Elvang:2013cua,Weinzierl:2016bus,Travaglini:2022uwo} for reviews on the always increasing number of techniques).
On the other hand, \textit{off-shell methods} are
very scarce (see e.g. \cite{Schubert:2001he} and references therein).
Off-shell amplitudes have in general a richer structure. Besides encoding the full on-shell information of a given process, off-shell results can be used in the computation of form factors (related to higher derivative terms in effective actions), in the study of quantum corrections of propagators and vertices of a theory, renormalization group analysis, and as building blocks for higher-loop corrections.

In between on-shell and off-shell methods, the Berends-Giele (BG)
currents \cite{Berends:1987me} stand out for their elegant and recursive
character. Simply put, BG currents represent tree-level amplitudes
with one off-shell leg, which are naturally interpreted as branches of higher-point trees. Even more interesting is the fact that the
BG prescription can be used to compute quantum effects. Indeed, one-loop
integrands are obtained by sewing tree-level amplitudes with two off-shell
legs. Following this idea, the matter contribution to one-loop amplitudes
in QCD was computed in \cite{Mahlon:1993si} by making an on-shell
matter leg of the BG current off-shell. This construction, however,
cannot be easily extended to gluon loops because of the gauge symmetry:
lifting the on-shell condition in gauge fields is a non-trivial task. In this case,
a concrete solution has only recently been found in \cite{Wu:2021exa},
following an extensive procedure using graphic rules in pure Yang-Mills. Numerical recursions for one-loop integrands are also known in the literature, see e.g. \cite{vanHameren:2009vq,Actis:2012qn}.

We would like to offer a fresh perspective on this subject, and show
how to obtain quantum \textit{off-shell} currents via classical equations
of motion \cite{footnote-intro}. The key ingredient is the perturbiner
\cite{Rosly:1996vr,Rosly:1997ap}, a well-known method to
recursively obtain Berends-Giele currents via formal multiparticle
solutions of the field equations. The perturbiner method has proven
to be a versatile technique to compute tree level amplitudes in different
theories \cite{Mafra:2015gia,Lee:2015upy,Mafra:2015vca,Mafra:2016ltu,Mafra:2016mcc,Mizera:2018jbh,Garozzo:2018uzj,Lopez-Arcos:2019hvg,Gomez:2020vat,Guillen:2021mwp,Gomez:2021shh,Ben-Shahar:2021doh,Escudero:2022zdz,Cho:2021nim}.
What we propose, instead, is to view it as a generator
of off-shell tree-level diagrams. This is achieved via multiparticle
ansatze that solve the interacting part of the field equations, while leaving single-particle states off-shell.
The standard notion of Berends-Giele currents is then generalised
to a fully off-shell version. In turn, we are able to define a consistent
sewing procedure to recursively generate one-loop off-shell currents,
dubbed \emph{one-loop pre-integrand}s. 

In this letter we focus on color-ordered theories, having the bi-adjoint
scalar and Yang-Mills theories as working examples. The sewing procedure
has to be supplied by a \emph{cyclic completion}, which is neatly
implemented using the multi-index structure of the one-loop pre-integrands.
It restores the cyclicity of the partial amplitude while solving the
combinatorial challenge of the Feynman approach. We start by illustrating
the construction in the bi-adjoint scalar theory with the derivation
of the off-shell recursion and the one-loop integrands. In Yang-Mills
we first perform the gauge-fixing of the action with the introduction
of Fadeev-Popov ghosts. We then extend the previous analysis to ghost
loop contributions, finally proposing the full $n$-point one-loop
integrand with off-shell external gluons.

\section{Bi-adjoint: off-shell recursion}

We will work here in $d$-dimensional Minkowski space with metric
$\eta_{\mu\nu}$ ($\mu,\nu=0,...,d-1$) and negative time signature.
We also use the shorthands $a\cdot b=a^{\mu}b^{\nu}\eta_{\mu\nu}$,
$k^{2}=k\cdot k$, and $\Box=\eta^{\mu\nu}\partial_{\mu}\partial_{\nu}$.

Consider a massless bi-adjoint scalar with cubic self-interaction
and classical equation of motion \cite{Cachazo:2013iea}
\begin{equation}
\Box\phi=\tfrac{1}{2}\llbracket\phi,\phi\rrbracket.\label{eq:BAEOM}
\end{equation}
Here we have $\phi=\phi_{a\tilde{a}}\:T^{a}\otimes\tilde{T}^{\tilde{a}}$,
where $a$ and $\tilde{a}$ are adjoint indices associated with two
different quadratic Lie algebras, with generators $T^{a}$ and $\tilde{T}^{\tilde{a}}$,
and
\begin{equation}
\llbracket T^{a}\otimes\tilde{T}^{\tilde{a}},T^{b}\otimes\tilde{T}^{\tilde{b}}\rrbracket=[T^{a},T^{b}]\otimes[\tilde{T}^{\tilde{a}},\tilde{T}^{\tilde{b}}].
\end{equation}

We are interested in building a generator of tree-level multiparticle
currents \textit{\`a la} Berends-Giele \cite{Berends:1987me}, but
through a modified perturbiner with \emph{off-shell} external legs.
To this end, we first look for a multiparticle solution to \eqref{eq:BAEOM}
of the form
\begin{equation}
\phi(x)=\sum_{P,Q}\Phi_{P\vert Q}\operatorname{e}^{\mathrm{i}k_{P}\cdot x}T^{a_{P}}\otimes\tilde{T}^{\tilde{a}_{Q}}.\label{eq:BAmultiparticle}
\end{equation}
The sum ranges over all words $P=p_{1}\cdots p_{n}$ and $Q=q_{1}\cdots q_{n}$ of length $\lvert P\rvert=\lvert Q\rvert=n$,
where $p_{i}$ and $q_{i}$ are single-particle labels. Furthermore,
we have $k_{P}=k_{p_{1}}+\cdots+k_{p_{n}}$, $T^{a_{P}}=T^{a_{1}}\cdots T^{a_{n}}$,
and $\tilde{T}^{\tilde{a}_{Q}}=\tilde{T}^{\tilde{a}_{1}}\cdots\tilde{T}^{\tilde{a}_{n}}$.
Inserting this back in \eqref{eq:BAEOM} leads to $k_{p}^{2}=0$ for
single-particle states and the following recursion relation,
\begin{equation}
\Phi_{P\vert Q}=\frac{1}{s_{P}}\sum_{P=RS}\sum_{Q=TU}[\Phi_{R\vert T}\Phi_{S\vert U}-(R\leftrightarrow S)],\label{eq:BArecursion}
\end{equation}
where $s_{P}=k_{P}^{2}$ are the Mandelstam variables. The sums over
$P=RS$ and $Q=TU$ denote deconcatenations of the word $P$ into
$R$ and $S$, and the word $Q$ into $T$ and $U$, respectively.
For example, for $P=$$ijk$ we have $(R,S)=(i,jk),(ij,k)$. By construction,
the coefficients $\Phi_{P\vert Q}$ automatically vanish unless the
words $P$ and $Q$ are related via permutation.

The multiparticle currents given by \eqref{eq:BArecursion} are identified
with Berends-Giele currents at tree level, and can then be used to
compute double-partial amplitudes \cite{Mafra:2016ltu}. On the other
hand, by dropping the on-shell condition for the single-particle states,
equation \eqref{eq:BArecursion} defines an off-shell recursion that
only solves the equation of motion \eqref{eq:BAEOM} at the multiparticle
level. For instance, this can be expressed as
\begin{equation}
\Box\phi-\tfrac{1}{2}\llbracket\phi,\phi\rrbracket=\sum_{p}k_{p}^{2}\Phi_{p\vert q}\operatorname{e}^{\mathrm{i}k_{p}\cdot x}T^{a_{p}}\otimes\tilde{T}^{\tilde{a}_{q}},\label{eq:BA-single}
\end{equation}
where the sum is taken over the single-particle states and $k_{p}^{2}\neq0$.
It is therefore fair to say that the recursion relations in \eqref{eq:BArecursion}
solve the interacting part of the biadjoint scalar theory while leaving
the single-particle states off-shell. In other words, they can be
interpreted as a generator of off-shell trees. Bearing this in mind,
we may refer to the coefficients $\Phi_{P\vert Q}$ as off-shell
Berends-Giele double-currents. 


We can then establish the connection between $\Phi_{P\vert Q}$ and
the off-shell scattering amplitudes in bi-adjoint scalar theory. This
is realized by a direct extrapolation of the Berends-Giele prescription,
such that the off-shell tree-level double-partial amplitudes are determined
through the formula 
\begin{equation}
m(Pn\vert Qn)=\lim_{k_{P}\to-k_{n}}s_{P}\Phi_{P\vert Q}\Phi_{n\vert n},\label{eq:BABGformula}
\end{equation}
where the limit enforces momentum conservation.

\section{Bi-adjoint: one-loop integrands}

Off-shell trees are the building blocks of loop amplitudes via a sewing
procedure. We will now show that the off-shell perturbiner expansion
leads to a simple algorithm for computing one-loop integrands in the
bi-adjoint scalar theory.

We start with the double-current $\Phi_{lP\vert lQ}$, in which the
single-particle label $l$ plays a special role. Using \eqref{eq:BArecursion},
such current can be explicitly expressed as
\begin{equation}
\begin{split}\Phi_{lP\vert lQ}= & \frac{1}{s_{lP}}(\Phi_{l\vert l}\Phi_{P\vert Q}+\sum_{P=RS}\sum_{Q=TU}\Phi_{lR\vert lT}\Phi_{S\vert U})\end{split}
\end{equation}
We can then factor out the single-particle polarization $\Phi_{l\vert l}$
on the right-hand side and recast $\Phi_{lP\vert lQ}$ as
\begin{equation}
\Phi_{lP\vert lQ}=\Phi_{l\vert l}\Lambda_{P\vert Q}(\ell),\label{eq:BAfactor}
\end{equation}
where $\ell^{\mu}\equiv k_{l}^{\mu}$, and
\begin{multline}
\Lambda_{P\vert Q}(\ell)=\frac{1}{(\ell+k_{P})^{2}}\\
\times[\Phi_{P\vert Q}+\sum_{P=RS}\sum_{Q=TU}\Lambda_{R\vert T}(\ell)\Phi_{S\vert U}].\label{eq:BA-BG-partialloop}
\end{multline}
 The double-current
$\Lambda_{P\vert Q}(\ell)$ is the fundamental ingredient for defining the one-loop
integrands. It needs but a small upgrade.

In order to see this, observe that \eqref{eq:BABGformula} and \eqref{eq:BAfactor}
yield 
\begin{equation}
m(lPn\vert lQn)=\lim_{k_{P}\to-\ell-k_{n}}(\ell+k_{P})^{2}\Phi_{l\vert l}\Phi_{n\vert n}\Lambda_{P\vert Q}(\ell).\label{eq:BA-sewing}
\end{equation}
The sewing procedure $\Phi_{l\vert l}\Phi_{n\vert n}\to1/\ell^{2}$,
with $k_{n}=-\ell$, leads to what looks like an onshell one-loop
integrand $I_{\textrm{1-loop}}(P\vert Q)\approx\Lambda_{P\vert Q}(\ell)$.
However, such an integrand is not cyclic in the words $P$ and $Q$, for the 
singling out of the leg $l$ has not been symmetrically  done. Fortunately,
the perturbiner framework enables a neat solution to this problem
via a \emph{cyclic completion} of the combinatorial sums defining
the recursion.

We then introduce the modified double-current,
\begin{multline}
\tilde{\Lambda}_{P\vert Q}(\ell)\equiv\frac{1}{(\ell+k_{P})^{2}}\\
\times[\Phi_{P\vert Q}+\tfrac{1}{2}\sum_{P=[RS]}\sum_{Q=[TU]}\Lambda_{R\vert T}(\ell)\Phi_{S\vert U}],\label{eq:BA-BG-fullloop}
\end{multline}
such that the one-loop integrand is expressed as
\begin{equation}
I_{\ell}^{\text{\ensuremath{1}-loop}}(P\vert Q)=\lim_{k_{P}\to0}\tilde{\Lambda}_{P\vert Q}(\ell).
\end{equation}

The words $P$ and $Q$ encode the (double) color ordering of the
one-loop integrand. The first term inside the square brackets only yields tadpole diagrams,
so it can be removed for convenience since their regularized contribution
vanishes. The cyclic completion affects the remaining terms, with
sums over $P=[RS]$ and $Q=[TU]$ consisting of all inequivalent cyclic
permutations of a given deconcatenation of $P$ and $Q$. The deconcatenations
of two cyclic permutations of the single-particle labels in $P$ and
$Q$ are equivalent if they lead to the same diagram contribution
on the one-loop integrand. As it turns out, they can be identified
via a simple rule.\footnote{During publication we had mistakenly upgraded this rule from our original submission, which we observed later was incorrect. The present version supersedes it.} For $P=[RS]$ we take the usual deconcatenation
of $P$ into the non-empty words $R$ and $S$, and then add $\lvert S\rvert-1$ cyclic
permutations in $P$. Take, for example, the word $P=1234$. The usual deconcatenations
are 
\begin{equation}
(R,S)=\textcolor{Blue}{(1,234)},\textcolor{Maroon}{(12,34)},\textcolor{PineGreen}{(123,4)}.\label{eq:ex1234}
\end{equation}
The operation $P=[RS]$ leads instead to
\begin{align}
(R,S)= & \textcolor{Blue}{(1,234)},\textcolor{Blue}{(2,341)},\textcolor{Blue}{(3,412)},\cancel{\textcolor{Blue}{(4,123)}},\nonumber \\
 & \textcolor{Maroon}{(12,34)},\textcolor{Maroon}{(23,41)},\cancel{\textcolor{Maroon}{(34,12)}},\cancel{\textcolor{Maroon}{(41,23)}},\label{eq:excyc1234}\\
 & \textcolor{PineGreen}{(123,4)},\cancel{\textcolor{PineGreen}{(234,1)}},\cancel{\textcolor{PineGreen}{(341,2)}},\cancel{\textcolor{PineGreen}{(412,3)}}.\nonumber 
\end{align}
The crossed permutations are not allowed. It is a straightforward
exercise to verify that this expression reproduces each of the diagrams
contributing to the one-loop off-shell integrand only once.

As an example, we will present the off-shell three-point one-loop
integrands. There are only two of them, namely $I_{\ell}^{\text{\ensuremath{1}-loop}}(123\vert123)$
and $I_{\ell}^{\text{\ensuremath{1}-loop}}(123\vert321)$. Using the
rules outlined above, we obtain
\begin{multline}
I_{\ell}^{\text{\ensuremath{1}-loop}}(123\vert123) +  I_{\ell}^{\text{\ensuremath{1}-loop}}(123\vert 321) \\= \frac{1}{\ell^2 (\ell+k_{1})^{2} (\ell+k_{1}+k_{2})^{2}},
\end{multline}
and
\begin{multline}
I_{\ell}^{\text{\ensuremath{1}-loop}}(123\vert321)= -\frac{1}{\ell^2} \bigg( \frac{1}{s_{12} (\ell+k_{1}+k_{2})^{2}}\\
+ \frac{1}{k_{1}^{2} (\ell+k_{1})^{2}} + \frac{1}{k_{2}^{2}(\ell+k_{2})^{2}}\bigg),
\end{multline}
with normalization $\Phi_{p\vert q}=\delta_{pq}$. These match the
results obtained in e.g. \cite{He:2015yua,Gomez:2017cpe}.

\section{Yang-Mills: off-shell recursion}

The main difference between the bi-adjoint model and the Yang-Mills
theory is the gauge symmetry. While at tree level, the gauge fixing
procedure is trivial, the loop construction naturally involves Faddeev-Popov
ghosts. The covariant gauge fixed action can then be cast as
\begin{multline}
S=\int d^{d}x\textrm{Tr}\{-\tfrac{1}{4}F_{\mu\nu}F^{\mu\nu}-\tfrac{1}{2\xi}(\partial^{\mu}A_{\mu})^{2}\\
+\partial^{\mu}b(\partial_{\mu}c-\mathrm{i}[A_{\mu},c])\},\label{eq:YM-gaugefixed}
\end{multline}
where $\xi$ is an arbitrary parameter. The gauge field $A_{\mu}$
and the ghost pair $(b,c)$ are Lie algebra valued. The field strength
is given by $F_{\mu\nu}=\partial_{\mu}A_{\nu}-\partial_{\nu}A_{\mu}-\mathrm{i}[A_{\mu},A_{\nu}]$,
with $A_{\mu}=A_{\mu}^{a}T^{a}$.

Our next step is to build the generator of tree level multiparticle
currents with off-shell external legs, just like in the bi-adjoint
case. Even though ghosts do not appear in classical configurations,
they will be treated for now as scalars with the wrong statistics.
The equations of motion derived from \eqref{eq:YM-gaugefixed} are
given by\begin{subequations}\label{eq:YM-eom-R}
\begin{align}
\Box A_{\mu} & =(1-1/\xi)\partial_{\mu}(\partial_{\nu}A^{\nu})-\mathrm{i}[A^{\nu},F_{\mu\nu}]\nonumber \\
 &\quad \: +\mathrm{i}\partial_{\nu}[A^{\nu},A_{\mu}]-\mathrm{i}\{\partial_{\mu}b,c\},\\
\Box b & =\mathrm{i}[A_{\mu},\partial^{\mu}b],\\
\Box c & =\mathrm{i}[A_{\mu},\partial^{\mu}c]+\mathrm{i}[\partial^{\mu}A_{\mu},c].
\end{align}
\end{subequations}

Note that the solutions of \eqref{eq:YM-eom-R} match classical Yang-Mills
solutions when $\partial^{\mu}A_{\mu}=b=c=0$. Multiparticle solutions
are obtained via the ansatz\begin{subequations}
\begin{eqnarray}
A_{\mu} & = & \sum_{P}\mathcal{A}_{P\mu}\operatorname{e}^{\mathrm{i}k_{P}\cdot x}T^{a_{P}},\label{eq:gauge-multiparticle}\\
b & = & \sum_{P}b_{P}\operatorname{e}^{\mathrm{i}k_{P}\cdot x}T^{a_{P}},\\
c & = & \sum_{P}c_{P}\operatorname{e}^{\mathrm{i}k_{P}\cdot x}T^{a_{P}}.
\end{eqnarray}
\end{subequations}The currents $\mathcal{A}_{P}^{\mu}$ reduce to
ordinary vector polarizations $\epsilon_{p}^{\mu}$ for one-lettered
words. The multiparticle ansatz can then be plugged back in \eqref{eq:YM-eom-R},
leading to the following recursions,\begin{subequations}

\begin{multline}
[\eta_{\mu\nu}s_{P}+\tfrac{(1-\xi)}{\xi}k_{P\mu}k_{P\nu}]\mathcal{A}_{P}^{\nu}=\sum_{P=QR}[k_{R\mu}b_{R}c_{Q}\\
\vphantom{\sum_{P=QR}}+\mathcal{A}_{Q}^{\nu}\mathcal{A}_{R}^{\rho}(k_{P\nu}\eta_{\mu\rho}+k_{R\nu}\eta_{\mu\rho}+k_{Q\mu}\eta_{\nu\rho})-(Q\leftrightarrow R)]\\
+\sum_{P=QRS}[\mathcal{A}_{Q}^{\nu}\mathcal{A}_{R}^{\rho}\mathcal{A}_{S}^{\sigma}(\eta_{\nu\sigma}\eta_{\mu\rho}-\eta_{\nu\rho}\eta_{\mu\sigma})+(Q\leftrightarrow S)],\label{eq:recursion-YM}
\end{multline}
\begin{equation}
b_{P}=-\tfrac{1}{s_P}\sum_{P=QR}[b_{Q}(k_{Q}\cdot\mathcal{A}_{R})-(Q\leftrightarrow R)],\label{eq:b-recursion}
\end{equation}
\begin{equation}
c_{P}=-\tfrac{1}{s_P}\sum_{P=QR}[c_{Q}(k_{P}\cdot\mathcal{A}_{R})-(Q\leftrightarrow R)],\label{eq:c-recursion}
\end{equation}
\end{subequations}The sum over $P=QRS$ denote deconcatenations
of the word $P$ into $Q$, $R$, and $S$.

Following the Faddeev-Popov procedure, all \emph{physical} single-particle
polarizations are transversal ($k_{p}\cdot\mathcal{A}_{p}=0$). Therefore,
$\mathcal{A}_{P}^{\mu}$ is identified with Berends-Giele currents
and can then be used to compute color ordered amplitudes via the usual
prescription \cite{Berends:1987me}. Just like in the bi-adjoint case,
we can determine a generator of off-shell trees by dropping the on-shell
condition for the single-particle states, and therefore solving the
equations of motion \eqref{eq:YM-eom-R} only at the multiparticle
level. For instance,
\begin{multline}
\partial^{\nu}F_{\mu\nu}-\tfrac{1}{\xi}\partial^{\mu}(\partial^{\nu}A_{\nu})-\mathrm{i}[A^{\nu},F_{\mu\nu}]\\
=\sum_{p}[k_{p}^{2}\mathcal{A}_{p\mu}+\tfrac{(1-\xi)}{\xi}k_{p\mu}(k_{p}\cdot\mathcal{A}_{p})]\operatorname{e}^{\mathrm{i}k_{p}\cdot x}T^{a_{p}},
\end{multline}
where the sum is now restricted to single-particle states. Interestingly, this is
enough to show that $\mathcal{A}_{P}^{\mu}$ still satisfies the shuffle
identity $\mathcal{A}_{Q\shuffle R}^{\mu}=0$, with $Q\shuffle R$ yielding the sum over all possible
shuffles between the words $Q$ and $R$. At tree level it
leads to the Kleiss-Kuijf relations \cite{Kleiss:1988ne}. At one-loop,
when we sew two legs and sum over specific cyclic permutations of
the remaining single-particle labels, the shuffle identity indirectly
leads to the Bern-Dixon-Dunbar-Kosower (BDDK) relations \cite{Bern:1994zx}. 

\section{Yang-Mills: gluon loop}

From now on we will work with $\xi=1$, which helps to implement the
recursion \eqref{eq:recursion-YM} through a scalar-like propagator
$1/s_{P}$. 

Towards the one-loop construction, let us consider the word $\hat{P}=lP$,
explicitly factorizing the single-particle label $l$, with associated
polarization $\epsilon_{l}^{\mu}$ and momentum $k_{l}^{\mu}$. In
this case we define $\mathcal{A}_{lP\mu}=\epsilon_{l}^{\nu}\mathcal{J}_{P\mu\nu}$,
with\begin{widetext}
\begin{align}
s_{lP}\mathcal{J}_{P\mu\nu} & =\mathcal{A}_{P\rho}[\delta_{\mu}^{\rho}(k_{lP}+k_{P})_{\nu}+\delta_{\nu}^{\rho}(k_{l}-k_{P})_{\mu}-\eta_{\mu\nu}(k_{l}+k_{lP})^{\rho}]+\sum_{P=QR}(2\delta_{\mu}^{\rho}\delta_{\nu}^{\sigma}-\delta_{\mu}^{\sigma}\delta_{\nu}^{\rho}-\eta_{\mu\nu}\eta^{\rho\sigma})\mathcal{A}_{Q\rho}\mathcal{A}_{R\sigma}\nonumber \\
 &\quad\: +\sum_{P=QR}[\delta_{\mu}^{\sigma}(k_{lP}+k_{R})^{\rho}-\delta_{\mu}^{\rho}(k_{lP}+k_{lQ})^{\sigma}+\eta^{\rho\sigma}(k_{l}+k_{Q}-k_{R})_{\mu}]\mathcal{J}_{Q\rho\nu}\mathcal{A}_{R\sigma}\label{eq:recursion-YM-J}\\
 &\quad \: +(2\delta_{\gamma}^{\rho}\delta_{\mu}^{\sigma}-\eta^{\rho\sigma}\eta_{\mu\gamma}-\delta_{\gamma}^{\sigma}\delta_{\mu}^{\rho})\sum_{P=QRS}\mathcal{J}_{Q\rho\nu}\mathcal{A}_{R\sigma}\mathcal{A}_{S}^{\gamma}.\nonumber 
\end{align}
\end{widetext}This is but a recasting of equation \eqref{eq:recursion-YM}
that singles out the particle $l$, though physically meaningful:
the current $\mathcal{J}_{P\mu\nu}$ is the one-loop pre-integrand
for a gluon loop.

Let us first examine the following object,
\begin{equation}
A(l,P,n)=\lim_{k_{lPn}\to 0}s_{lP}(\epsilon_{l}^{\nu}\mathcal{J}_{P\mu\nu})\epsilon_{n}^{\mu},
\end{equation}
where $\epsilon_{n}^{\mu}$ is the polarization of an off-shell leg
with momentum $k_{n}^{\mu}$. The analogy with \eqref{eq:BA-sewing}
is clear. The sewing procedure is simply $\epsilon_{l}^{\mu}\epsilon_{n}^{\nu}\to\eta^{\mu\nu}/k_{l}^{2}$,
with $k_{l}=-k_{n}^{\mu}$, yeilding a look-alike one-loop integrand
$I^{\text{\ensuremath{1}-loop}}(P)\approx\eta^{\mu\nu}\text{\ensuremath{\mathcal{J}_{P\mu\nu}}}$
for a single-trace color-ordered correlator. Once more, the issue
with this construction is that the current $\mathcal{J}_{P\mu\nu}$
is no longer cyclic in the word $P$, so its cyclic completion has
to be introduced by hand.

As in the bi-adjoint case, we take the loop momentum to be $\ell^{\mu}\boldsymbol{\equiv}k_{l}^{\mu}$,
with $k_{P}^{\mu}=0$. If we explicitly remove tadpole contributions,
given by the first line in \eqref{eq:recursion-YM-J}, the one-loop
integrand can be cast as
\begin{equation}
I_{\textrm{gluon}}^{\text{\ensuremath{1}-loop}}(P;\ell)\equiv\eta^{\mu\nu}\text{\ensuremath{\tilde{\mathcal{J}}_{P\mu\nu}}}(\ell),\label{eq:one-loop-offshell-integrand}
\end{equation}
with
\begin{multline}
\tilde{\mathcal{J}}_{P\mu\nu}=\frac{1}{2\ell^{2}}\sum_{P=[QR]}\mathcal{J}_{Q\rho\mu}\mathcal{A}_{R\sigma}\\
\times[\delta_{\nu}^{\sigma}(k_{R}+\ell)^{\rho}-\delta_{\nu}^{\rho}(k_{Q}+2\ell)^{\sigma}+\eta^{\rho\sigma}(2k_{Q}+\ell)_{\nu}]\\
+\frac{1}{\ell^{2}}\sum_{P=[QRS]}\mathcal{J}_{Q\rho\mu}\mathcal{A}_{R\sigma}\mathcal{A}_{S\gamma}\\
\times(2\delta_{\nu}^{\sigma}\eta^{\gamma\rho}-\eta^{\sigma\gamma}\delta_{\nu}^{\rho}-\delta_{\nu}^{\gamma}\eta^{\rho\sigma}).\label{eq:YM-integrand-current}
\end{multline}
The sum over $P=[QR]$ is analogous to the bi-adjoint case. In particular the factor of $1/2$ accounts for the equivalent contributions mentioned after equation  \eqref{eq:excyc1234}. The sum
over $P=[QRS]$ is simpler, related to quartic vertices in Yang-Mills:
the deconcatenations of all cyclic permutations in $P$ are inequivalent
and have to be included. 

Because of the color structure, as in the bi-adjoint construction,
\eqref{eq:one-loop-offshell-integrand} reproduces each of the diagrams
contributing to the one-loop off-shell integrand without repetitions.
The only subtlety missed by \eqref{eq:one-loop-offshell-integrand}
is the automorphism of the two-point integrand with $P=12$ and $P=21$,
which has to be taken into account separately. It corresponds to the
color-stripped one-loop bubble diagram with external moment $k_{2}^{\mu}=-k_{1}^{\mu}=k^{\mu}$.
From \eqref{eq:recursion-YM-J} and \eqref{eq:YM-integrand-current},
we obtain
\begin{equation}
I_{\textrm{gluon}}^{\text{\ensuremath{1}-loop}}(12;\ell)=\frac{1}{2}\frac{1}{\ell^{2}(\ell-k)^{2}}N_{\mu\nu}\epsilon_{1}^{\mu}\epsilon_{2}^{\nu},
\end{equation}
with
\begin{multline}
N^{\mu\nu}=[\delta_{\rho}^{\mu}(\ell-2k)_{\sigma}+\eta_{\rho\sigma}(k-2\ell)^{\mu}+\delta_{\sigma}^{\mu}(\ell+k)_{\rho}]\\
\times[\eta^{\nu\sigma}(\ell+k)^{\rho}+\eta^{\rho\sigma}(k-2\ell)^{\nu}+\eta^{\nu\rho}(\ell-2k)^{\sigma}],
\end{multline}
matching the textbook computation using Feynman diagrams. The extra
factor $1/2$  cancels the automorphism over counting.

We have also checked that equation \eqref{eq:one-loop-offshell-integrand}
reproduces the three- and four-point one-loop integrands obtained
using Feynman rules. These are of course only consistency checks, since our construction goes far beyond. Just like at tree level, the recursive implementation
of \eqref{eq:recursion-YM} and \eqref{eq:recursion-YM-J} greatly
simplify the involved computational work, and can be easily implemented
via commonly available software for symbolic computation.

\section{Yang-Mills: ghost loop}

The steps for the definition of a ghost loop are very similar to the
ones taken before. In this case, we have to consider off-shell multiparticle
currents with one ghost external leg labeled by $l$, either $b_{l}$
or $c_{l}$, such that $b_{lP}=b_{l}\mathcal{B}_{P}$ and $c_{lP}=c_{l}\mathcal{C}_{P}$, 
with\begin{subequations}\label{eq:ghost-currents-factorized}

\begin{align}
\mathcal{B}_{P} & =-\frac{1}{s_{lP}}[(k_{l}\cdot\mathcal{A}_{P})+\sum_{P=QR}\mathcal{B}_{Q}(k_{lQ}\cdot\mathcal{A}_{R})],\\
\mathcal{C}_{P} & =-\frac{1}{s_{lP}}[(k_{lP}\cdot\mathcal{A}_{P})+\sum_{P=QR}\mathcal{C}_{Q}(k_{lP}\cdot\mathcal{A}_{R})].
\end{align}
\end{subequations}The currents $\mathcal{B}_{P}$ and $\mathcal{C}_{P}$
involve only gluons, since the ghost polarization has been explicitly
stripped off. 

We then define the one-loop integrands\begin{subequations}
\begin{eqnarray}
\tilde{\mathcal{B}}_{P}(\ell) & = & -\frac{1}{\ell^{2}}\sum_{P=[QR]}\mathcal{B}_{Q}(k_{lQ}\cdot\mathcal{A}_{R}),\\
\tilde{\mathcal{C}}_{P}(\ell) & = & -\frac{1}{\ell^{2}}\sum_{P=[QR]}\mathcal{C}_{Q}(k_{l}\cdot\mathcal{A}_{R})],
\end{eqnarray}
\end{subequations}where $k_{P}=0$, and the tadpole contributions
have been removed. The cyclic completion is being enforced in the
sums following the same recipe of the pure gluon case, yielding the full diagrammatic expansion without redundant contributions. Note that $\tilde{\mathcal{B}}_{P}=\tilde{\mathcal{C}}_{P}$ when $(k_{Q}\cdot\mathcal{A}_{Q})=0$, i.e., when the external gluons are on-shell.


The full one-loop integrand with off-shell external gluons is
\begin{equation}
I_{\ell}^{\text{\ensuremath{1}-loop}}(P)=I_{\textrm{gluon}}^{\text{\ensuremath{1}-loop}}(P;\ell)-\tilde{\mathcal{C}}_{P}(\ell),\label{eq:full-integrand}
\end{equation}
where the ghost contribution appears with a minus sign (fermionic
loop).

As an example, we present the one-loop gluon self-energy, 
\begin{equation}
\Sigma(k)=\int d^{d}\ell(\tfrac{1}{2}\eta^{\mu\nu}\text{\ensuremath{\tilde{\mathcal{J}}_{12\mu\nu}}}-\tilde{\mathcal{C}}_{12})\equiv\epsilon_{1}^{\mu}\epsilon_{2}^{\nu}\Pi_{\mu\nu}(k),
\end{equation}
where $k\equiv k_{2}=-k_{1}$, $\Pi_{\mu\nu}=\Pi(k^{2})(k^{2}\eta_{\mu\nu}-k_{\mu}k_{\nu})$,
and 
\begin{multline}
\Pi(k^{2})=\tfrac{\mathrm{i}}{2}\pi^{d/2}(-k^{2})^{d/2-2}(5-7d)\\
\times\frac{\Gamma(d/2)^{2}\Gamma(1-d/2)}{\Gamma(d)}.
\end{multline}
The loop integral was performed via an analytic continuation from
$k^{2}<0$ and the dimensional regularization is implicit. The poles
of $\Gamma(1-d/2)$ encode infrared as well as ultraviolet divergences.


\section{Final remarks}

We have established here a robust framework for computing one-loop off-shell
integrands in color-ordered theories from classical equations of motion.
The main ingredient in this construction is the observation that multiparticle
solutions can be viewed as a generator of off-shell trees in a given
field theory. Because of the color structure, it is straightforward
to identify the cyclic completion necessary to obtain the full one-loop
integrand.

By construction, the one-loop integrands in \eqref{eq:BA-BG-fullloop}
for the bi-adjoint scalar and in \eqref{eq:full-integrand} for Yang-Mills
lead to single-trace partial amplitudes, which we denote by $A_{n;0}$
(with $n=|P|$). Therefore, the full one-loop amplitudes in these
theories are more conveniently represented by the Del Duca-Dixon-Maltoni
color decomposition \cite{DelDuca:1999rs}. For instance, in Yang-Mills
we have
\begin{equation}
A_{\textrm{tot}}^{\text{\ensuremath{1}-loop}}=\sum_{\sigma\in S_{n-1}/\mathcal{R}}c_{n}(\sigma)A_{n;0}(\sigma_{1},...,\sigma_{n}),\label{eq:YM-full-loop-amplitude}
\end{equation}
where $\sigma$ is the color order (i.e., the word $P$ in our formulas),
and $c_{n}(\sigma)$ is the color basis defined by nested commutators
of the group generators,
\begin{equation}
c_{n}(\sigma)\equiv\textrm{Tr}(T^{a}[T^{\sigma_{1}},[...,[T^{\sigma_{n-1}},[T^{\sigma_{n}},T^{a}]]...]]),
\end{equation}
with shorthand $T^{\sigma}=T^{a_{\sigma}}$. In the sum, $S_{n-1}$
denotes permutations of $(n-1)$ legs, and $\mathcal{R}$ denotes
reflection. Although \eqref{eq:YM-full-loop-amplitude} is expanded
in a single trace basis, it encodes also the double-trace (non-planar)
contributions, which can be determined via the BDDK relations (see
e.g. \cite{Du:2014uua} for a nice summary of different color decompositions).

The recursive character of the one-loop pre-integrands \eqref{eq:BA-BG-partialloop},
\eqref{eq:recursion-YM-J}, and \eqref{eq:ghost-currents-factorized},
represent an objective simplification over the traditional diagrammatic
approach, since they can be algorithmically implemented without extra
effort. More than that, their present form enable a transparent identification
of the corresponding diagrams. For example, the external leg bubbles
can be read off from \eqref{eq:YM-integrand-current} through the
currents $\mathcal{A}_{R\mu}$ with $|R|=|P|-1$. And finally, the
one-loop pre-integrands can be directly used as building blocks for
higher-loop integrands via different sewings, since all external legs
are off-shell. 

An interesting feature in our proposal is that \textit{a priori} no Lagrangian is required. Therefore we can compute the one-loop off-shell scattering of field theories that are known only at the level of equations of motion. For example, six-dimensional $\mathcal{N}=(2,0)$ superconformal field theory, which is supposed to describe the low energy limit of $M5$-branes \cite{Claus:1997cq,Howe:1997fb}. See also \cite{Lyakhovich:2005mk,Lyakhovich:2006sc,Lyakhovich:2007cw} for more details on the quantization of non-Lagrangian theories. More generally, our results can be applied to a variety of colored theories, including non-linear sigma model, Chern-Simons, super Yang-Mills, etc. For
gauge theories coupled to matter, like QCD, our method becomes
slightly more involved since we lose the rigidity of the color structure.
Preliminary results on this will be reported in a different
work. Perhaps more noteworthy is the fact we can generalize the results
of \cite{Gomez:2021shh} to off-shell graviton trees, including ghosts,
therefore introducing a recursive tool for computing one-loop off-shell integrands
in Einstein gravity \cite{GLLQ:2022}.


\begin{acknowledgments}
We would like to thank Subhroneel Chakrabarti, Arthur Lipstein, Sebastian Mizera, and Christian Schubert  for useful discussions
and feedback on the manuscript. We would also like to thank Sebastian
Mizera for sharing his perturbiner notebook with us. Finally, we would like to thank the anonymous referees for comments and suggestions. HG is supported by the Royal Society via a PDRA grant.
\end{acknowledgments}

\end{document}